\documentclass[sigconf]{aamas} 

\usepackage{balance} 
\usepackage{multirow}

\setcopyright{ifaamas}
\acmConference[AAMAS '25]{Proc.\@ of the 24th International Conference
on Autonomous Agents and Multiagent Systems (AAMAS 2025)}{May 19 -- 23, 2025}
{Detroit, Michigan, USA}{A.~El~Fallah~Seghrouchni, Y.~Vorobeychik, S.~Das, A.~Nowe (eds.)}
\copyrightyear{2025}
\acmYear{2025}
\acmDOI{}
\acmPrice{}
\acmISBN{}

    \title[AAMAS-2025 Formatting Instructions]{Revisiting Communication Efficiency in Multi-Agent Reinforcement Learning from the Dimensional Analysis Perspective}

\author{Chuxiong Sun}
\affiliation{
  \institution{National Key Laboratory of Space Integrated Information System, Institute of Software Chinese Academy of Sciences}
  \city{Beijing}
  \country{China}}
\email{chuxiong2016@iscas.ac.cn}
\authornote{Equal contribution.}

\author{Peng He}
\affiliation{
  \institution{Beijing University of Posts and Telecommunications}
  \city{Beijing}
  \country{China}}
\email{hepeng123@bupt.edu.cn}
\authornotemark[1]

\author{Rui Wang}
\affiliation{
  \institution{National Key Laboratory of Space Integrated Information System, Institute of Software Chinese Academy of Sciences \\
  State Key Laboratory of Intelligent Game
  }
  \city{Beijing}
  \country{China}}
\email{wangrui@iscas.ac.cn}

\author{Changwen Zheng}
\affiliation{
  \institution{National Key Laboratory of Space Integrated Information System, Institute of Software Chinese Academy of Sciences}
  \city{Beijing}
  \country{China}}
\email{changwen@iscas.ac.cn}
\authornote{Corresponding author.}

\begin{abstract}
In this work, we introduce a novel perspective—dimensional analysis—to address the challenge of communication efficiency in Multi-Agent Reinforcement Learning (MARL). Our findings reveal that simply optimizing the content and timing of communication at sending end is insufficient to fully resolve communication efficiency issues. Even after applying optimized and gated messages, dimensional redundancy and confounders still persist in the integrated message embeddings at receiving end, which negatively impact communication quality and decision-making. To address these challenges, we propose Dimensional Rational Multi-Agent Communication (DRMAC), designed to mitigate both dimensional redundancy and confounders in MARL. DRMAC incorporates a redundancy-reduction regularization term to encourage the decoupling of information across dimensions within the learned representations of integrated messages. Additionally, we introduce a dimensional mask that dynamically adjusts gradient weights during training to eliminate the influence of decision-irrelevant dimensions. We evaluate DRMAC across a diverse set of multi-agent tasks, demonstrating its superior performance over existing state-of-the-art methods in complex scenarios. Furthermore, the plug-and-play nature of DRMAC’s key modules highlights its generalizable performance, serving as a valuable complement rather than a replacement for existing multi-agent communication strategies.
 
\end{abstract}


\newcommand{\BibTeX}{\rm B\kern-.05em{\sc i\kern-.025em b}\kern-.08em\TeX}

\begin{document}


\pagestyle{fancy}
\fancyhead{}


\maketitle 


\section{Introduction}
\label{sec:intro}
Reinforcement Learning (RL) has achieved significant milestones across a wide array of complex real-world domains, including Game AI \cite{osband2016deep,silver2017mastering,silver2018general,vinyals2019grandmaster}, Robotics \cite{robotics}, and Autonomous Driving \cite{leurent2018survey}. However, in cooperative multi-agent settings, distinct challenges emerge. One of the most prominent is the issue of partial observability, where agents are restricted to their local observations, lacking a comprehensive view of the entire environment. Further complicating the situation, Multi-Agent Reinforcement Learning (MARL) faces the problem of non-stationarity, as the environment's dynamics constantly shift from the perspective of individual agents, thereby introducing significant complexity into the learning process. 
Multi-agent communication offers a promising solution to these challenges by enabling agents to gain a more comprehensive understanding of their surroundings through shared information. This approach not only stabilizes the learning process but also fosters coordinated actions among agents, ultimately improving overall system performance.

\begin{figure}
    \centering
    \includegraphics[width=1\linewidth]{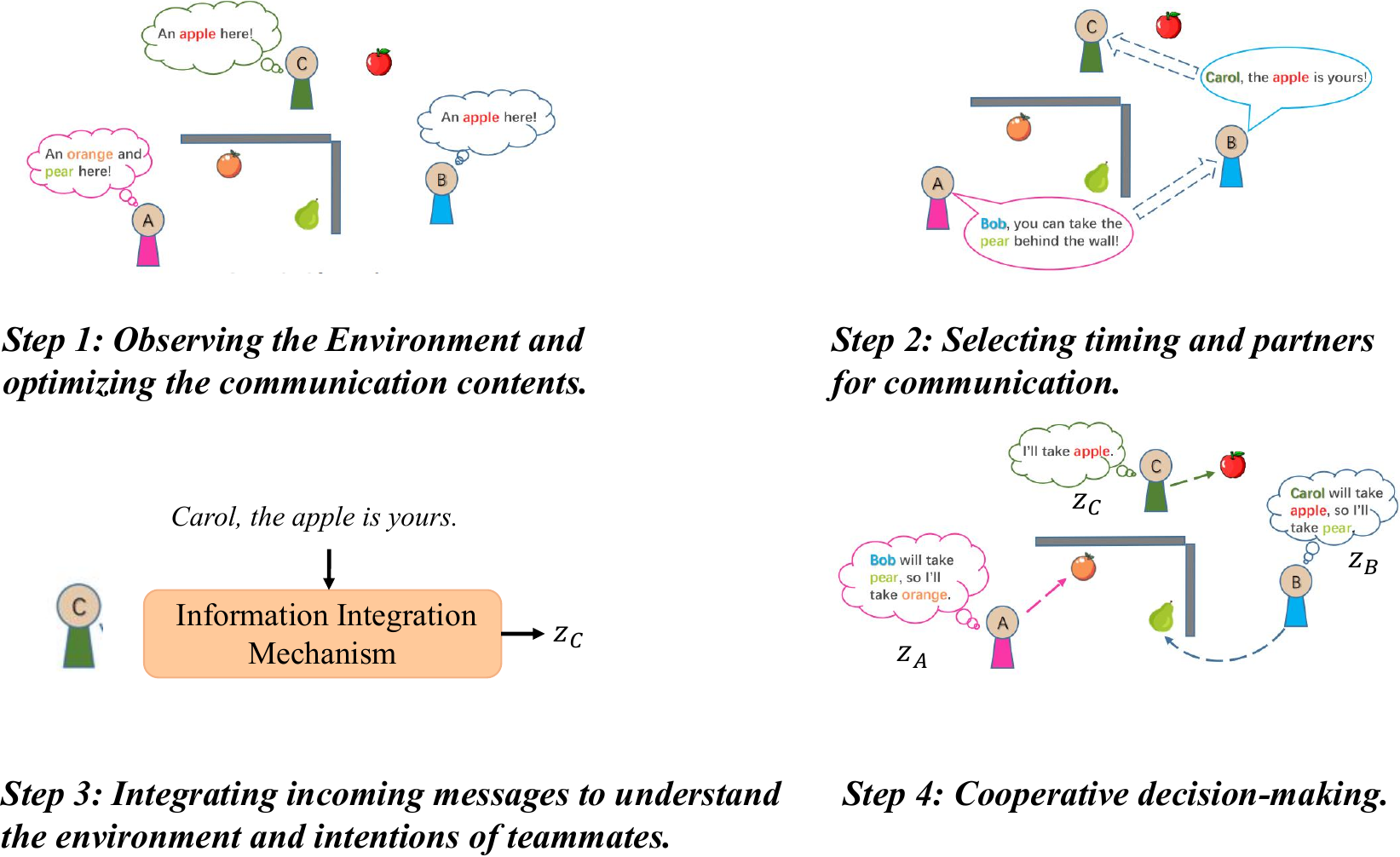}
    \caption{Existing communication methods typically focus on Step 1 and Step 2, optimizing the content, timing, and selection of partners for communication, with the aim of enhancing efficiency at the message level. However, in this work, we observe that even after these optimizations, dimensional redundancy and confounders still exist in the integrated message embeddings at Step 3. To further improve communication efficiency, we introduce a novel perspective—dimensional analysis—as a complement to existing communication methods.}
    \label{fig:commparadigm}
\end{figure}

\begin{figure*}
    \centering
    \includegraphics[width=0.9 \linewidth]{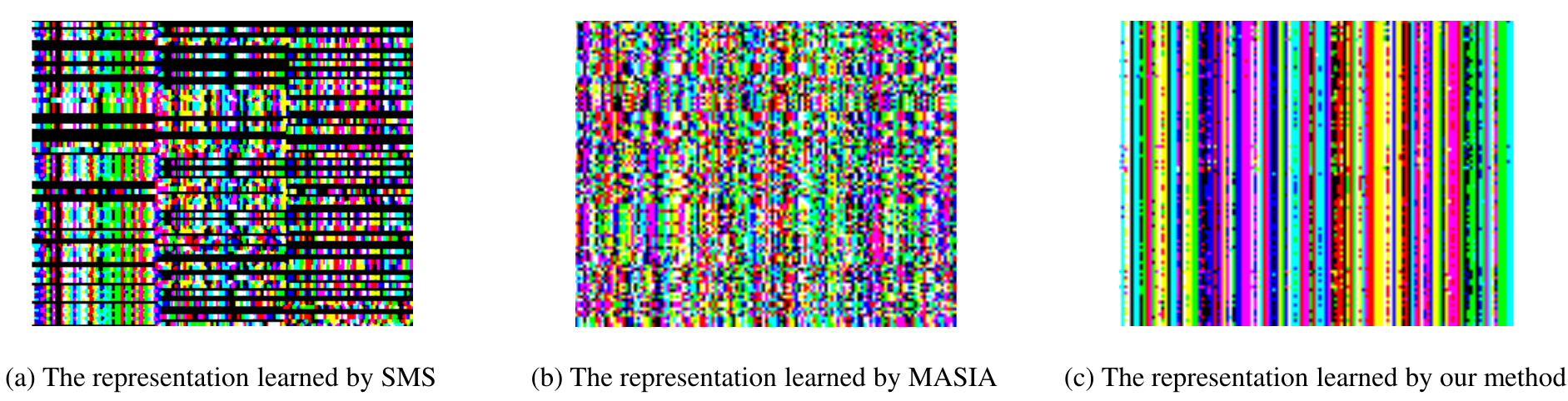}
    \caption{The visualizations depict the representations learned by SMS\cite{sms}, MASIA\cite{masia} and our method on a challenging SMAC task, 1o\_2r\_vs\_4r. The learned features are projected into an RGB color image, where distinct colors indicate different feature types. The horizontal axis corresponds to feature dimensions, while the vertical axis represents samples from different trajectories. Greater color contrast signifies lower similarity between feature dimensions. These plots illustrate the similarity between dimensional features within a batch. In contrast to existing approaches, our method employs a redundancy-reduction technique that efficiently decouples the received information into distinct dimensions. Each of these dimensions represents a unique part of the information's entropy, ensuring that the resulting representation is both more informative and less redundant.}
    \label{fig:redundancy}
\end{figure*}

\begin{figure*}
    \centering
    \includegraphics[width=0.9\linewidth]{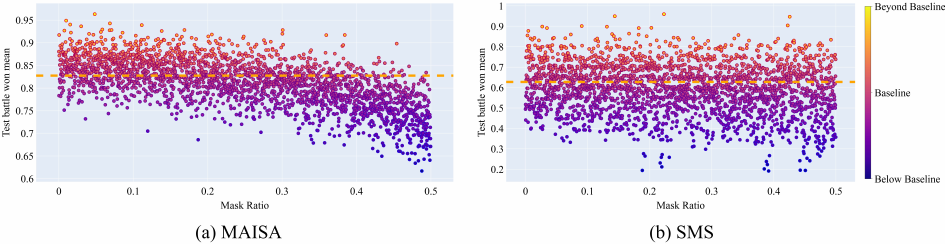}
    \caption{Experimental scatter plots were generated by MASIA and SMS with randomly masked dimensions on a challenging SMAC task, 1o\_2r\_vs\_4r. The Baseline and the red dashed lines indicate the performance achieved by the unmasked representation of MASIA and SMS. Each point represents an independent experimental result, obtained by applying a specific mask rate to the original representation at the dimensional level. Notably, the original representation remains unchanged throughout the experiments. These results demonstrate the pervasive presence of dimensional confounders in the process of multi-agent communication.}
    \label{fig:confounder}
\end{figure*}

An essential challenge in multi-agent communication is optimizing communication efficiency. A critical observation is that only a subset of agents provide meaningful insights, while overwhelming the network with redundant information can be not only ineffective but also detrimental to learning\cite{jiang2018learning}. Consequently, existing approaches have focused on refining both the \textbf{content} and \textbf{timing} of communication \cite{commnet111,ic3net,schednet,ndq,tmc,maic}. These methods typically employ regularization techniques to reduce message redundancy or assess the relevance of observed information to determine optimal timing and partner selection for communication, thus minimizing unnecessary exchanges. We categorize these approaches as message-level de-redundancy strategies.

Although message-level de-redundancy strategies have made significant strides, state-of-the-art methods continue to face challenges related to communication inefficiency.  Hence, in this work, we revisit the multi-agent communication paradigm. As illustrated in Fig.\ref{fig:commparadigm}, once agents optimize the content and timing of their communication and share the messages with teammates, the receiving agents employ an encoder to generate embeddings that assimilate these messages. The essence of multi-agent communication lies in utilizing these embeddings to enhance agents' understanding of an uncertain environment, ultimately leading to more refined decision-making policies. Hence, in this work, we conduct dimensional analysis on these embeddings, identifying \textbf{dimensional redundancy} and \textbf{dimensional confounders} as the fundamental issues that hinder efficient communication.

Specifically, from an information theory perspective, each dimension encapsulates a portion of the representation's information entropy. Dimensional redundancy occurs when multiple dimensions contain overlapping information entropy. To demonstrate the existence of such dimensional redundancy, we visualize the integrated representations produced by both MASIA\cite{masia} and SMS\cite{sms}. MASIA represents a full-communication approach, while SMS employs a gating mechanism to filter messages at sending end. As illustrated in Fig.\ref{fig:redundancy}, despite messages being gated and optimized at the sending end, substantial dimensional redundancy still persists at the receiving end. Therefore, it remains essential for the receiving agents to perform global filtering of all received information to further enhance communication efficiency.

The dimensional confounder refers to a set of dimensions containing "harmful" information, which ultimately degrades the model's performance. To demonstrate the existence of dimensional confounders, we conduct motivating experiments using both MASIA and SMS. Specifically, we randomly mask several dimensions of message representation at receiving end by setting their values to zero, and subsequently evaluate the performance of the masked message representations. The experimental results are presented in Fig.\ref{fig:confounder}, where each point represents the outcome for several evaluated trajectories at a given mask ratio. Interestingly, representations of received messages with certain masked dimensions can achieve better performance compared to their unmasked counterparts. This empirical evidence supports our hypothesis regarding dimensional confounders, indicating that even after agents have filtered information during transmission, some decision-irrelevant information remains embedded in the representations integrated by the receiving agents.

With this vision in mind, we propose Dimensional Rational Multi-Agent Communication (DRMAC) to address both dimensional redundancy and confounders in MARL. Specifically, we propose a dimensional redundancy reduction regularization term that drives the cross-correlation matrix of twin message embeddings towards an identity matrix. The twin message embeddings are generated by taking the received messages as input along with a perturbed encoder of the same input, thereby creating two correlated views for contrastive learning.  This simple yet powerful approach effectively reduces dimensional redundancy in the integrated message embeddings.  
To further combat the influence of dimensional confounders, we introduce a dimensional mask, which enhances the gradient effect of decision-relevant dimensions while reducing the influence of confounding dimensions during training. The learnable dimensional mask is optimized through a meta-learning paradigm, with the goal of enhancing the performance of masked representations on MARL tasks. Consequently, the dimensional mask continuously adjusts the weights of various dimensions based on their gradient contributions, encouraging the model to focus on acquiring specific decision-relevant information. This approach effectively addresses communication inefficiencies, particularly in complex tasks involving numerous agents and communication messages. We rigorously evaluated DRMAC across multiple MARL environments, including Hallway and SMAC. When compared to leading multi-agent communication strategies such as TarMac \cite{tarmac}, MAIC \cite{maic}, SMS \cite{sms}, and MASIA \cite{masia}, DRMAC consistently demonstrated superior performance and efficiency, highlighting its robustness and effectiveness in diverse settings. 
Our contributions can be summarized as follows:
\begin{itemize} 
\item We revisit the existing paradigm of multi-agent communication and identify that optimizing the timing and content of messages at the receiving end alone is insufficient to address communication efficiency issues. Therefore, we introduce a novel perspective—dimensional analysis—to enhance the performance of multi-agent communication in MARL. 
\item We propose DRMAC, which incorporates a redundancy reduction regularization term and an Information Selective Network (ISN) to address the challenges of dimensional redundancy and dimensional confounders, respectively. 
\item Empirically, we demonstrate that DRMAC not only facilitates more effective message integration but also significantly enhances communication efficiency, thereby addressing a critical research gap in MARL.
\end{itemize}

\section{Related Works}

\subsection{Multi-Agent Communication}
Multi-agent communication has emerged as an indispensable component in MARL\cite{vdn,qmix1,wang2021qplex,mappo1,maddpg}. 
Research in this domain has primarily concentrated on three fundamental questions:  

\emph{Determining the optimal content of communication (what to communicate).} 
CommNet \cite{commnet111}, as a pioneering work in this domain, enabled agents to learn continuous messages for effective communication. Building on the foundation laid by CommNet, numerous approaches have been proposed to further refine the message learning process. VBC \cite{vbc} focuses on filtering out noise while preserving valuable information by restricting message variance. TMC \cite{tmc} introduces regularizers to minimize temporal redundancy in messages, thereby enhancing efficiency. NDQ \cite{ndq} leverages information-theoretic regularizers to generate expressive yet concise messages. MAIC \cite{maic} advances communication by allowing agents to customize messages for specific recipients, thereby promoting tailored and more effective information exchange.

\emph{Deciding appropriate timing and partners for information exchange (when and whom to communicate).}
To improve communication efficiency, methods like IC3Net \cite{ic3net} and ATOC \cite{schednet} have employed gating networks to eliminate unnecessary communication links. Similarly, SchedNet \cite{schednet} and IMMAC \cite{immac} utilize heuristic mechanisms to evaluate the significance of observations, thereby gating non-essential communication and reducing overhead. Additionally, techniques such as MAGIC \cite{magic1}, I2C \cite{I2C1}, and SMS \cite{sms} have been developed to identify the most appropriate recipients for messages. These methods focus on assessing the impact of shared information on the recipients' decision-making, ensuring that communication is directed where it can most effectively enhance decision-making outcomes.

\emph{Integrating incoming messages and making decisions (how to utilize received information).} 
TarMAC \cite{tarmac} has explored how agents can effectively assimilate crucial information from an abundance of raw messages. MASIA \cite{masia} represents the pioneering effort to introduce self-supervised representation learning into multi-agent communication. Specifically, MASIA proposed two self-supervised auxiliary tasks to enhance the efficiency of information integration. However, this paper primarily focuses on addressing the issues of dimensional redundancy and dimensional confounders in multi-agent communication representation learning.

\subsection{Self-supervised learning}
Self-supervised learning approaches \cite{2020Debiased, 2020Hard, 2020Tsai, 2020Bootstrap, 2020WhiteningErmolov, 2021Barlow, 2021Vikas, 2021Tete} have achieved significant breakthroughs in unsupervised representation learning. Specifically, CMC \cite{Tian2019Contrastive} and AMDIM \cite{2019Philip} leverage contrastive learning techniques on multi-view datasets, thereby enhancing representation quality by integrating multiple views. SimCLR \cite{chen2020simple} and MoCo \cite{2020Kaiming} utilize large batch sizes or memory banks to expand the pool of negative samples, facilitating the learning of more discriminative and robust representations. SwAV \cite{2020Mathilde}, on the other hand, compares cluster assignments across different views instead of directly comparing features, thereby leveraging multiple views to improve learning efficiency. Methods such as BYOL \cite{2020Bootstrap} and Barlow Twins \cite{2021Barlow} have also identified a key challenge—insufficient self-supervision can lead to feature collapse, where learned representations degenerate into trivial solutions. Nonetheless, these benchmark self-supervised learning frameworks benefit from well-designed architectures that incorporate sufficient self-supervision, resulting in improved discriminative capabilities of the learned representations. 
Recently, these advantages have been extended to Visual Reinforcement Learning (VRL)\cite{curl,drq,rad,ucb-rad}, where contrastive learning is employed to derive informative and discriminative low-dimensional representations from high-dimensional visual inputs. Additionally, MA2CL\cite{ma2cl}, a MARL method, has adopted self-supervised learning to encourage the learned representations to be both temporally and agent-level predictive by reconstructing masked agent observations within the latent space. Consequently, a growing number of RL approaches are integrating self-supervised learning techniques to develop more refined state representations. 
Unlike prior methods that primarily aim to enhance state representations by combining self-supervised learning with RL, DRMAC models the information integration challenge in multi-agent communication as a representation learning problem. Specifically, DRMAC seeks to enhance communication efficiency by reducing redundancy and mitigating confounders in the integrated representations, thereby significantly improving the quality of communication among agents.

\section{Background}
In this work, we focus on fully cooperative multi-agent tasks, characterized by partial observability and necessity for inter-agent communication. These tasks are modeled as Decentralized Partially Observable Markov Decision Processes (Dec-POMDPs) \cite{decpomdp}, represented by the tuple $G = (N, S, O, A, \mathbb{O}, P, R, \gamma, M)$. In this formulation, $N \equiv \{1, \ldots, n\}$ denotes the set of agents, $S$ represents the global states, $O$ describes the observations available to each agent, $A$ signifies the set of available actions, $\mathbb{O}$ is the observation function mapping states to observations, $P$ is the transition function illustrating the dynamics of the environment, $R$ is the reward function dependent on the global states and joint actions of the agents, $\gamma$ is the discount factor, and $M$ specifies the set of messages that can be communicated among the agents. 
At each time-step $t$, each agent $i \in N$ has access to its own observation $o_i^t \in O$ determined by the observation function $\mathbb{O}(o_i^t | s_t)$. Additionally, each agent can receive messages $c_i^t = \sum_{j \neq i}m_j^t$ from teammates $j \in N$. Utilizing both the observed and received information, agents then make local decisions. As each agent selects an action, the joint action $a_t$ results in a shared reward $r_t = R(s_t, a_t)$ and transitions the system to the next state $s_{t+1}$ according to the transition function $P(s_{t+1}|s_t,a_t)$. The objective for all agents is to collaboratively develop a joint policy $\pi$ to maximize the discounted cumulative return $\sum_{t=0}^{T}\gamma^{t}r_{t}$.

\section{Methods}
\begin{figure*}
    \centering
    \includegraphics[width=1\linewidth]{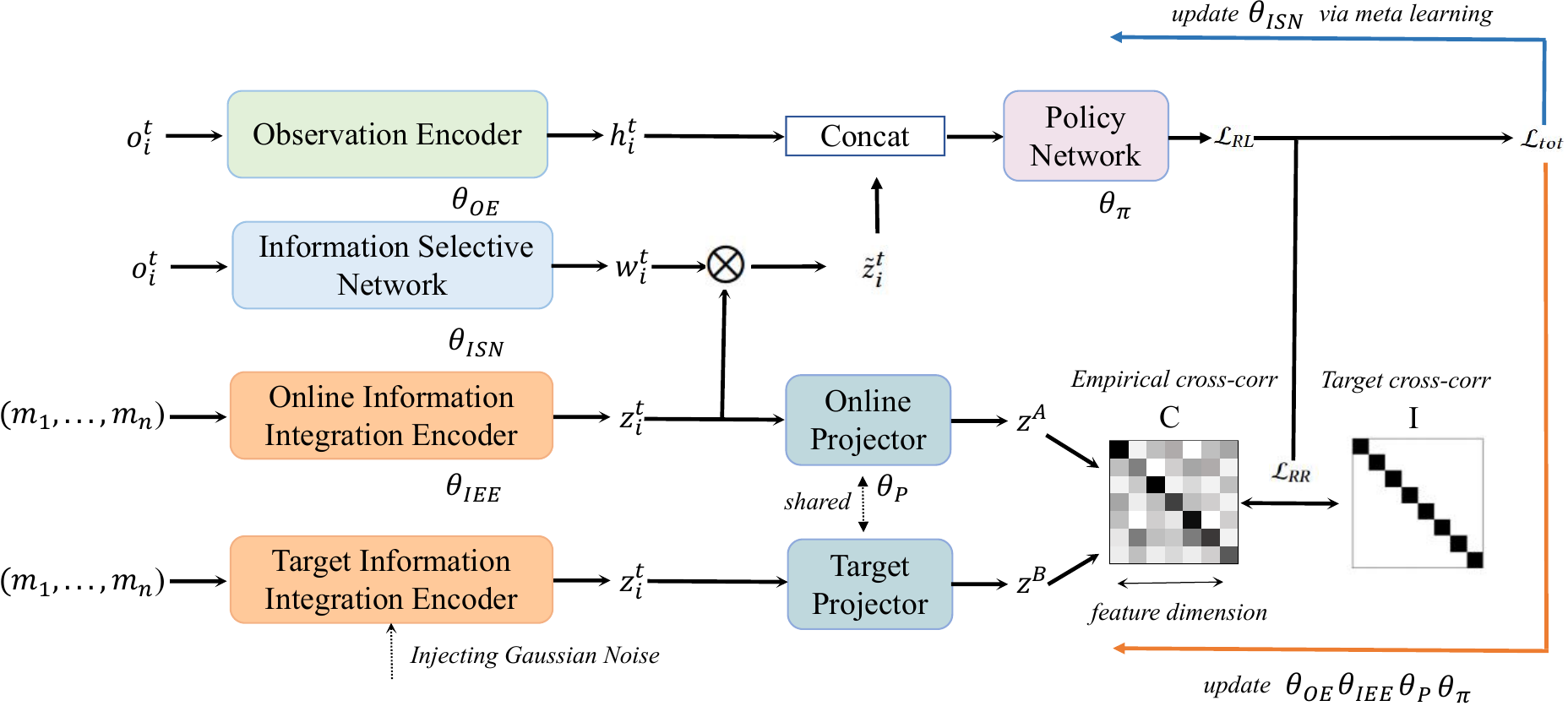}
    \caption{The overview of our proposed DRMAC.}
    \label{fig:framework}
\end{figure*}

As illustrated in Fig. \ref{fig:framework}, upon receiving messages $(m_1^t,...,m_n^t)$, DRMAC incorporates an Information Integration Encoder(IIE) to aggregate the information shared by agents into a common representation $z_t$. Our goal is to reduce dimensional redundancy and confounders in this representation, thereby ensuring that each agent receives sufficient and relevant information to make effective decisions. The key features of DRMAC are highlighted as follows:
\begin{itemize}
    \item \textbf{Redundancy Reduction Regularization}: DRMAC’s integration network at the receiving end is optimized using a redundancy reduction regularization term, which aims to decouple the received information into distinct dimensions, with each dimension encapsulating a unique subset of the representation's information entropy.
    \item \textbf{Information Selective Network (ISN)}: After encoding a decoupled representation, the ISN learns a dimensional mask that adjusts the gradient weights of each dimension according to their contributions to the decision-making process. This allows the policy network to focus on relevant and significant dimensional information while effectively ignoring dimensional confounders.
\end{itemize}
We provide the details of the redundancy reduction mechanism in \textbf{Section \ref{subsec:rr}}, and the specifics of ISN in \textbf{Section\ref{subsec:isn}}. Additionally, a comprehensive description of the training process is outlined in \textbf{Section\ref{subsec:training}}.

\subsection{Dimensional Redundancy Reduction for Integrated Message Embeddings}
\label{subsec:rr}
As illustrated in Fig. \ref{fig:redundancy}, despite optimizing communication content and timing at message level, multi-agent communication methods still face the issue of dimensional redundancy—where multiple dimensions in the integrated message embeddings contain overlapping information entropy. To address this challenge, we draw inspiration from a neuroscience approach, specifically redundancy reduction \cite{barlow1961possible}, which hypothesizes that the goal of sensory processing is to recode highly redundant sensory inputs into a factorial code, consisting of statistically independent components. Building on its successful application in self-supervised learning \cite{2021Barlow}, we apply dimensional redundancy reduction to integrated message embeddings to mitigate overlapping information and improve communication efficiency. Concretely, for agent $i$, the received messages are first fed into the Information Integration Encoder to obtain its representation $z_i^t$. To obtain multiple views of the message embeddings, we duplicate the parameters of the IIE and add Gaussian noise to the copied version. We then encode the identical set of received messages using both the original IIE and the noise-added IIE, resulting in two embeddings, $z_i^t$ and ${z_i^t}'$. Essentially, this process serves as a form of data augmentation. Furthermore, previous works \cite{chen2020simple} have experimentally demonstrated that introducing a nonlinear layer before the contrastive loss can significantly improve performance. Therefore, we also include a non-linear projection layer. This layer is implemented as an MLP network. To simplify notations, we utilize $z^A$ and $z^B$ to denote the two embeddings after online projector and target projector respectively. The redundancy reduction objective is formulated as follows:
\begin{equation}
    \mathcal{L}_{RR} \triangleq 
\underbrace{\sum_i \left(1 - C_{ii}\right)^2}_{\text{invariance term}} 
\quad +  \underbrace{\lambda\sum_i \sum_{j \neq i} C_{ij}^2}_{\text{redundancy reduction term}}
\end{equation}
where $\lambda$ is a hyper-parameter, and $C$ is computed between the outputs of the online and target projectors:
\begin{equation}
    C_{ij} \triangleq 
\frac{\sum_b z_{b,i}^A z_{b,j}^B}
     {\sqrt{\sum_b \left(z_{b,i}^A\right)^2} \sqrt{\sum_b \left(z_{b,j}^B\right)^2}}
\end{equation}
where $b$ indexes batch samples, and $i$, $j$ index the vector dimensions of the network outputs. The matrix $C$ is a square matrix with a size equal to the dimensionality of the network's output, and its values range between -1 (indicating perfect anti-correlation) and 1 (indicating perfect correlation).

Intuitively, the invariance term of the objective aims to set the diagonal elements of the cross-correlation matrix to 1, thereby ensuring that the embeddings remain invariant to applied distortions. Meanwhile, the redundancy reduction term seeks to bring the off-diagonal elements of the cross-correlation matrix to 0, effectively decorrelating the different components of the embedding vector. This decorrelation minimizes redundancy among the output units, ensuring that each unit captures unique, non-overlapping information in the embeddings of the received messages. With the centralized process enhanced by $\mathcal{L}_{RR}$, the online Information Integration Encoder can efficiently integrate incoming messages and produce decoupled representations. It is important to note that this process is only applied during the training phase and is not utilized during execution.

\begin{figure*}
    \centering
    \includegraphics[width=0.7\linewidth]{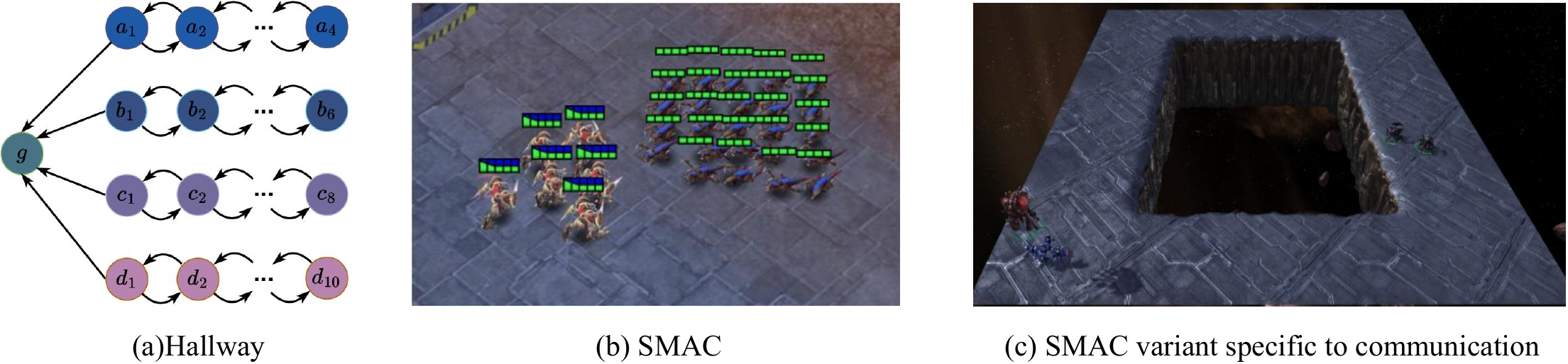}
    \caption{Multiple environments considered in our experiments.}
    \label{fig:env}
\end{figure*}

\subsection{Information Selective Network: Addressing Dimensional Confounders}
\label{subsec:isn}
To address the issue of dimensional confounders, we use the Information Selective Network (ISN) to construct a learnable dimensional mask $\omega = \left\{ \omega_k \mid k \in \llbracket 1, D \rrbracket \right\}$. Considering that, in the CTDE framework, receiving agents operate in different situations and fulfill distinct roles within the team, the dimensional confounders in $z$ may vary for each agent. Therefore, we design the ISN to take each agent's observation as input and produce a dimensional mask specific to that agent. The ISN is implemented as a Multi-Layer Perceptron (MLP) with a sigmoid activation function at the output layer, ensuring that each dimension of $\omega_k$ is bounded between 0 and 1. The dimensional mask is then used to assign a weight to each dimension of the representation.
\begin{equation}
    \tilde{z}_i^t = z_i^t \otimes \omega_i^t
\end{equation}
where $\tilde{z}_i^t$ denotes the masked representation, $\omega_i^t$ represents the dimensional mask for agent $i$ at time-step $t$ and $\otimes$ is an element-wise Hadamard product function. By applying element-wise multiplication to $z_i^t$, each agent derives a distinct representation tailored to its specific needs. This process enables each agent to extract only the information from the decoupled representation that is most relevant to its decision-making context, while ignoring dimensions that serve as confounders. Then, the masked representation, representing receiving agent's understanding about received messages are concated with its own local observations to make the local decisions, $a_i^t=\pi_{\theta_{\pi}}(o_i^t, \tilde{z}_i^t)$. This tailored approach ensures that each agent’s representation is aligned with its unique requirements, thereby enhancing the quality of decision-making and improving overall efficiency.

\subsection{Overall Training Process}
\label{subsec:training}
The ISN is designed to explore dimensional confounders within multiple observed information, specifically tailoring the extracted representation to meet the diverse needs of local decision-making for multiple agents. The primary challenge here lies in the dynamic nature of the MARL environment and the variability in communication needs at different stages of a mission. In contrast to static environments, the significance of certain information dimensions can change dramatically, requiring ISN to continuously adapt and maintain its effectiveness. This complexity makes it difficult to solve the problem through traditional optimization methods, which often rely on first-order gradients and are insufficient in scenarios characterized by rapidly changing conditions and diverse agent requirements. 
Hence, we utilize meta-learning to update the ISN. Meta-learning provides the ISN with the ability to learn how to adapt quickly to new situations by leveraging past experiences. This approach enables the ISN to continuously adjust to the dynamic communication needs of agents, efficiently dealing with the ever-changing importance of different information dimensions. 

In our training framework, it is crucial to emphasize that only the parameters of $\theta_{ISN}$ are refined through the meta-learning process, while the rest of the system parameters are updated via standard first-order gradient methods. Specifically, in the first regular training step, we concentrate on optimizing the parameter set $\theta = (\theta_{OE}, \theta_{IEE}, \theta_{P}, \theta_{\pi})$ by jointly minimizing both the redundancy reduction loss and the RL loss, as described by:
\begin{equation}
    \arg\min_{\theta}\mathcal{L}_{tot}(\theta, \theta_{ISN}), 
\end{equation}
where $\mathcal{L}_{tot}(\theta, \theta_{ISN}) = \mathcal{L}_{RL} + \beta \mathcal{L}_{RR}$, and $\beta$ is a weighting coefficient to balance the RL objective with the redundancy reduction objective. It is worth noting that our DRMAC framework is compatible with any MARL training algorithm and multi-agent communication strategy. The experimental results presented in \textbf{Section\ref{sec；gen}} substantiate this assertion, 

In the second meta-learning-based step, $\theta_{ISN}$ is updated using a meta-learning approach that leverages second-order gradient information\cite{meta}. This method is crucial for adapting $\theta_{ISN}$ to properly recognize the importance of each information dimension. The update involves calculating the gradient of $\theta_{ISN}$ based on the combined performance metric $\mathcal{L}_{RL}$, and is formulated as follows:
\begin{equation}
    \arg\min_{\theta_{ISN}}\mathcal{L}_{RL}(\theta_{trial}, \theta_{ISN}),
\label{updatemeta}
\end{equation}
where $\theta_{trial} = (\theta_{OE}^{trial}, \theta_{IEE}^{trial}, \theta_{P}^{trial}, \theta_{\pi}^{trial})$ are the trial weights obtained after a single gradient update on $\mathcal{L}_{RL}$. This update of trial weights is computed as:
\begin{equation}
\theta_{trial}=\theta-\ell_\theta\nabla_\theta\mathcal{L}_{RL},  
\label{trailcompute}
\end{equation}
with $\ell_\theta$ representing the learning rate. Notably, during the calculation of these trial weights, back-propagation of gradients is excluded to ensure computational efficiency. Consequently, $\theta_{ISN}$ is refined through second-order gradient optimization based on $\theta$. This process allows ISN to be continuously fine-tuned by considering the gradient contributions from $\mathcal{L}_{RL}$, enabling it to effectively discern the relevance of each information dimension and adapt accordingly. By using this meta-learning-based approach, ISN dynamically identifies and focuses on critical dimensions, enhancing decision-making quality and ensuring communication efficiency.

We train the dimensional mask $\omega$ solely based on RL performance for two key reasons. First, dimensional redundancy reduction acts as a regularization term that imposes strong constraints. Second, redundancy reduction is independent of downstream decision-making tasks, while our primary goal is to use the masked representation to improve agents' understanding of environmental uncertainty and enhance cooperative decision-making. By focusing on RL performance to train $\omega$, we enable agents to effectively identify and mitigate dimensional confounders that are irrelevant to cooperative decision-making.

\begin{figure*}
    \centering
    \includegraphics[width=1\linewidth]{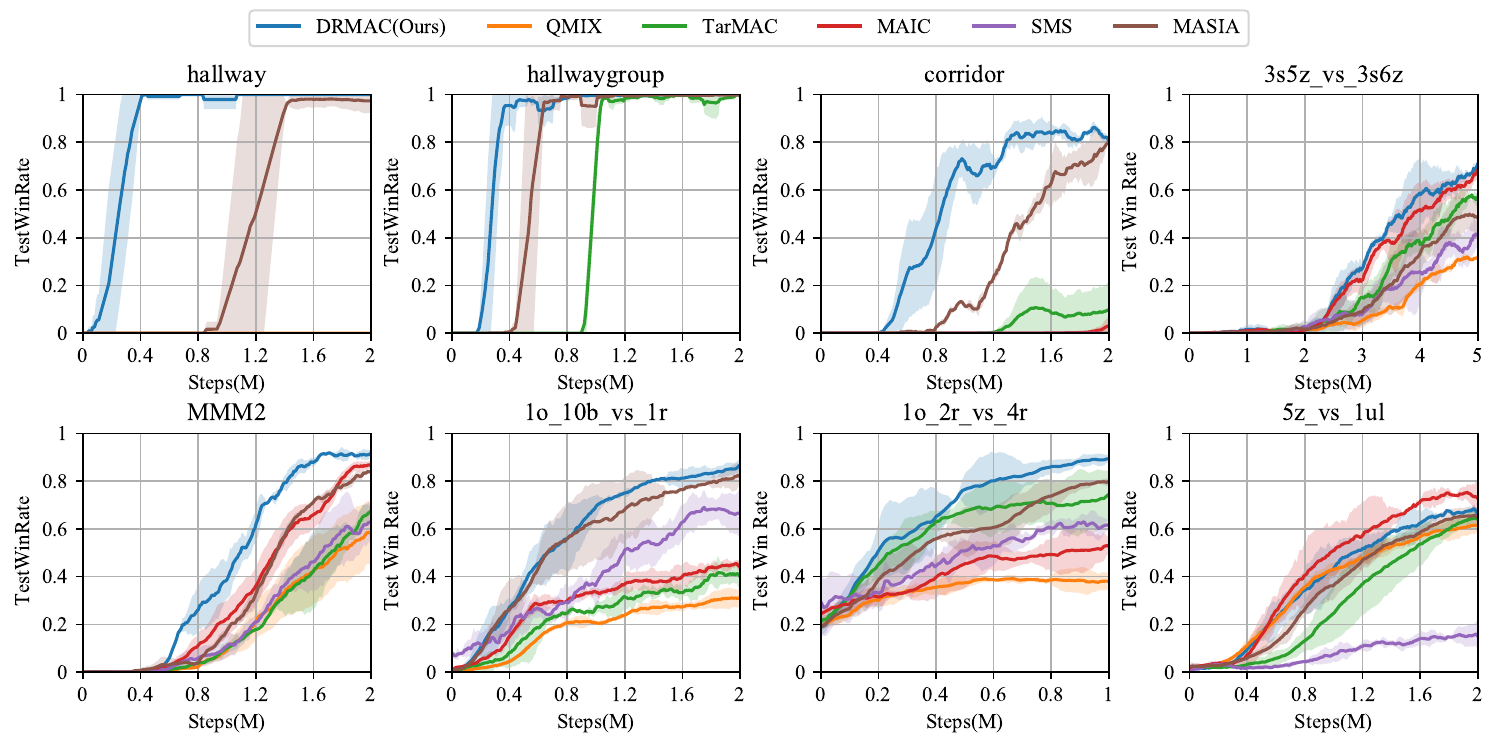}
    \caption{Performance on multiple tasks.}
    \label{fig:performance}
\end{figure*}

\section{Experiments}
In this section, we present our experimental design, which is meticulously structured to address three fundamental research questions:

\begin{itemize} 
\item \textbf{RQ1.} How does the performance of DRMAC compare to that of state-of-the-art communication methods? 
\item \textbf{RQ2.} Which specific components of DRMAC are crucial to its overall performance? 
\item \textbf{RQ3.} Can DRMAC be effectively integrated with various existing communication baselines to enhance their communication efficiency? 
\end{itemize}

\subsection{Setup}
\textbf{Baselines} To address RQ1, we compare DRMAC against a diverse set of baselines, including a strong communication-free baseline, QMIX \cite{qmix1}, as well as several state-of-the-art communication methods, such as MASIA \cite{masia}, MAIC \cite{maic}, SMS \cite{sms}, and TarMAC \cite{tarmac}. In our experiments, the implementation of each baseline method strictly follows the official repositories, with all hyperparameters left at their original, best-performing configurations to ensure fair and consistent comparison.

\textbf{Environments} As depicted in Fig. \ref{fig:env}, we conduct a comprehensive evaluation of DRMAC across two well-known cooperative multi-agent environments, encompassing a total of eight tasks that vary in the number of agents, difficulty levels, and task types.We begin with the Hallway environment \cite{ndq}, which provides a relatively simple setup based on multiple Markov chains. In this environment, agents are randomly positioned within different chains and must collaborate to reach a goal state simultaneously, operating under conditions of partial observability. To further assess cooperative performance, we introduce an improved variant named HallwayGroup, featuring diverse agents grouped into different cohorts, each with distinct requirements for reaching their respective goal states at different times. 
SMAC \cite{smac} is derived from the well-known real-time strategy game StarCraft \uppercase\expandafter{\romannumeral2}, focusing on micromanagement challenges where each unit is independently controlled by an agent making decisions under partial observability. To increase task difficulty, we selected three "superhard" tasks, which are particularly challenging due to their complex opponent compositions and dynamic game states. Additionally, to further stress the difficulties inherent in multi-agent communication, we incorporated three more tasks from the setup described in \cite{ndq}. These tasks not only restrict the agents' vision range but also immerse them in highly intricate maps, characterized by maze-like terrain or unpredictable spawning dynamics of both allied and enemy units. This combination of limited visibility and complex environments provides a rigorous assessment of DRMAC's ability to manage coordination under demanding conditions.

\textbf{Implementation.} Our DRMAC implementation is built upon PyMARL and QMIX, utilizing a basic full communication strategy at the sending end. For more details about implementation, please refer to Appendix.\ref{app:implementation}.

\subsection{Performance(RQ1)}
Our evaluation begins with a comparative analysis of the learning curves of DRMAC against several baseline methods across diverse environments. This comparison aims to assess DRMAC's overall cooperative performance and communication efficiency. As shown in Fig. \ref{fig:performance}, DRMAC consistently outperforms the baselines in almost all environments, demonstrating its robust capabilities across a wide range of cooperative multi-agent tasks of varying difficulties and scales. This superior performance can be attributed to DRMAC's ability to effectively counteract dimensional redundancy and confounders, allowing it to consistently identify the most relevant dimensions in the integrated message embeddings at the receiving end. It is worth noting that, across nearly all environments, the baselines employing communication significantly outperform the communication-free baseline. This result suggests that the selected environments and tasks are inherently communication-intensive and challenging, further reinforcing DRMAC's superior performance in handling such complex scenarios. To gain a deeper understanding of the intrinsic mechanisms behind DRMAC's performance, we conduct a more detailed analysis of specific tasks. In SMAC, five of the six selected tasks, excluding 5z\_vs\_1ul, involve controlling diverse types of agents. For these tasks, existing methods stress the importance of designing communication architectures (e.g., SMS) and tailoring communication content (e.g., MAIC) to align with the specific capabilities of each unit. However, we observe that both MAIC and SMS do not perform consistently well across these five tasks. For instance, while MAIC shows strong performance in tasks like 3s5z\_vs\_3s6z and MMM2, it underperforms in other heterogeneous scenarios. In contrast, DRMAC, as well as MASIA, consistently demonstrates superior performance across all tasks. 
This observation further supports our analysis that optimizing communication filtering and content at the sending end is only the first step toward enhancing communication efficiency. Even after optimizing the communication content and timing, receiving agents must still carefully integrate the received information to form embeddings that are truly useful for decision-making. Moreover, the consistently superior performance of DRMAC underscores the importance of addressing dimensional redundancy and confounders, highlighting DRMAC's effectiveness in overcoming these challenges.

\subsection{Ablation(RQ2)}
\begin{figure}
    \centering
    \includegraphics[width=0.56\linewidth]{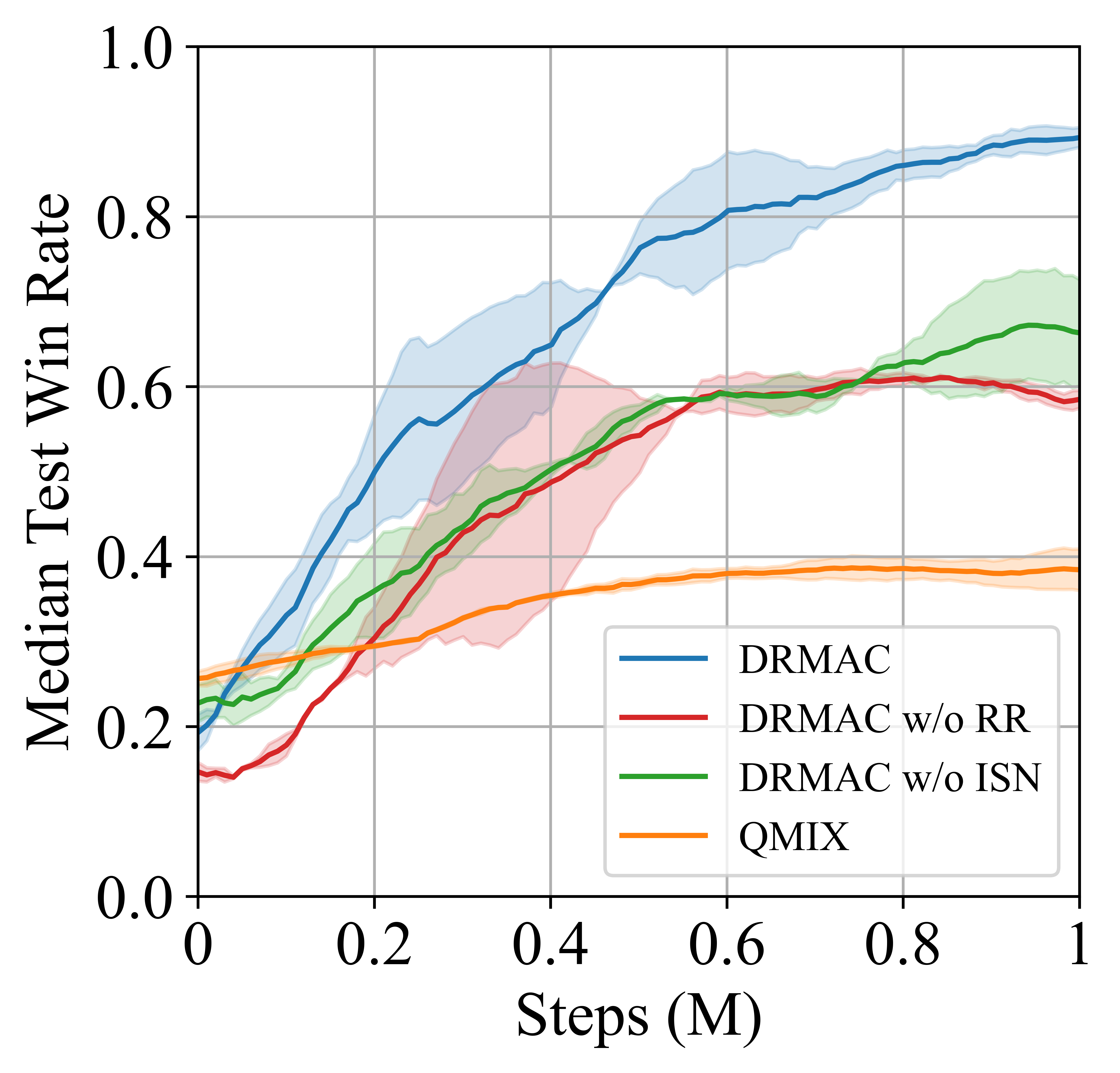}
    \caption{Performance of different DRMAC variants.}
    \label{fig:ablation}
\end{figure}

We conduct several ablation studies, primarily focusing on the scenario $1o\_2r\_vs\_4r$,  to evaluate the specific contributions of different components in DRMAC. Below is a breakdown of the configurations examined: 
\begin{itemize}
    \item \textbf{DRMAC}:This refers to the complete method proposed in our work, incorporating all components.
    \item \textbf{QMIX}:Serving as a baseline, QMIX represents the core functionality without the enhancements introduced by DRMAC.
    \item \textbf{DRMAC w/o RR}:This is a variant of DRMAC where the redundancy reduction techniques are removed, allowing us to assess the impact of dimensional redundancy reduction.
    \item  \textbf{DRMAC w/o ISN}:This variant of DRMAC omits ISN, treating each dimension of the representation as equally important. This allows us to assess the significance of mitigating dimensional confounders.
\end{itemize}
As illustrated in Figure \ref{fig:ablation}, the performance of both DRMAC w/o RR and DRMAC w/o ISN is suboptimal. The poor performance of DRMAC w/o ISN aligns with our expectations, as redundancy reduction primarily aims to decouple the received information, which is not directly relevant to decision-making performance. However, the underperformance of DRMAC w/o RR further underscores the significance of the proposed redundancy reduction regularization. This result indicates that relying solely on ISN is insufficient to achieve state-of-the-art performance. We attribute this to the presence of dimensional redundancy, which forces ISN to identify agent-specific dimensional confounders from high-dimensional redundant representations, thereby increasing the complexity of training ISN. In summary, our approach of extracting dimensional confounders from redundancy-reduced representations is mutually reinforcing, with both components being essential for optimal performance.

\subsection{Generalizability(RQ3)}
\label{sec；gen}
\begin{figure}
    \centering
    \includegraphics[width=0.6\linewidth]{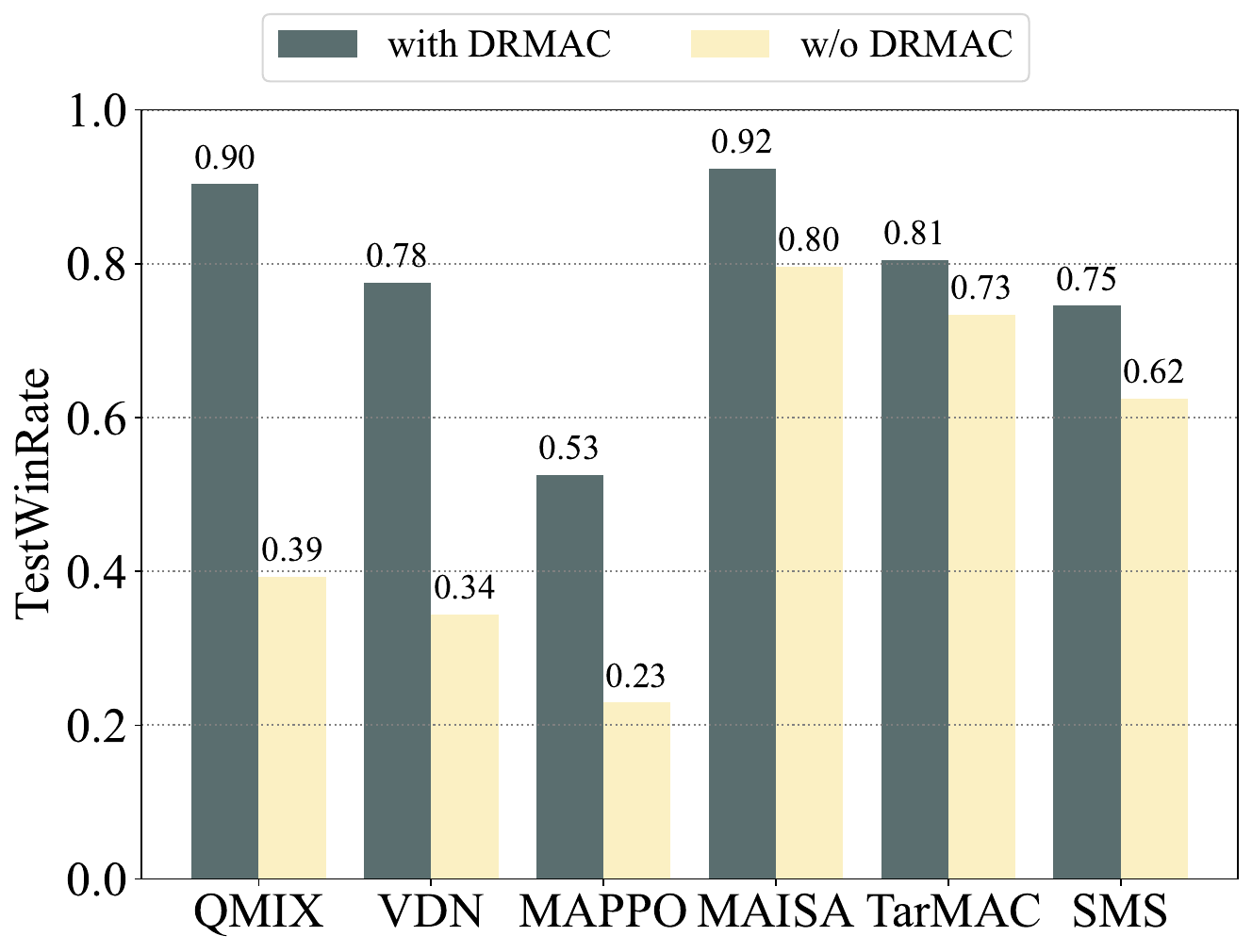}
    \caption{Performance of DRMAC integrated with various MARL baselines.}
    \label{fig:generation}
\end{figure}

We begin by evaluating the generalizability of DRMAC by combine it with different communication-free MARL methods, such as VDN\cite{vdn}, QMIX\cite{qmix1} and MAPPO\cite{mappo1}. As illustrated in Fig.\ref{fig:generation}, DRMAC consistently delivers superior performance across all these baselines, often with a significant improvement margin. This positive performance enhancement underscores the broad applicability and effectiveness of DRMAC in addressing dimensional redundancy and confounders, ultimately leading to improved communication efficiency and cooperative performance in MARL.

To further demonstrate the generalizability of DRMAC, we implement it as a plug-and-play module into several state-of-the-art multi-agent communication methods, specifically MASIA\cite{masia}, SMS\cite{sms}, and TarMAC\cite{tarmac}. Among these methods, MASIA and TarMAC are fully communication-based, whereas SMS incorporates message filtering at the sender side. As illustrated in Fig. \ref{fig:generation}, we observe that DRMAC, as a plug-and-play module, consistently enhances the communication performance of these state-of-the-art methods. This result further validates the presence of dimensional redundancy and confounders, as well as their negative impact on communication efficiency. Moreover, these findings highlight the significance and effectiveness of studying the dimensional rationale at the receiving end, regardless of whether message-level filtering has already been applied at the sender side. In other words, applying DRMAC to conduct dimensional analysis on the representations after the received messages are encoded can serve as a complementary, rather than a replacement, to existing approaches.

\section{Conclusion}
In this work, we addressed the challenge of communication efficiency in MARL by introducing a novel perspective—dimensional analysis—to complement existing methods. Our experimental results demonstrate that optimizing the content and timing of communication alone is insufficient to fully address the issue of communication efficiency, as dimensional redundancy and confounders continue to persist. To overcome these challenges, we proposed DRMAC, a method designed to mitigate both dimensional redundancy and confounders in MARL. Specifically, we introduced a redundancy-reduction regularization term that encourages the dimensions of the learned message representations to model decoupled information. Furthermore, we employed a dimensional mask to adjust the gradient weights of different dimensions during training, thereby effectively eliminating the influence of decision-irrelevant dimensions. Our approach not only enhances the efficiency of communication but also leads to more robust and effective decision-making in MARL.



\bibliographystyle{ACM-Reference-Format} 
\bibliography{sample}


\newpage
\appendix
\onecolumn

\section{Implementation Details}
\label{app:implementation}
\subsection{Network architecture}
\begin{table}[h]
\centering
\begin{tabular}{ll}
  \toprule
  \textbf{Module}  & Architecture \\
  \midrule
  \multirow{5}{*}{Information Integration Encoder}       & Q:Linear(obs\_dim, 16)\\
                                    & K:Linear(obs\_dim, 16)\\
                                    & V:Linear(obs\_dim, 32)\\ 
                                    & RNN(32, 32)\\
                                    & Linear(32,128)\\\hline
  Observation Encoder                 & Linear(obs\_dim,64)\\\hline
  
  Projector                 & 
                                Linear(128,512)\\
                                &
                                Linear(512,512)\\\hline

  Policy Network                & Linear(128+64,32)\\
                                & RNN(32, 32)\\
                                & Linear(32, n\_action)\\
  
  \bottomrule
  Information Selective Network                &Linear(obs\_dim,128)\\
  \bottomrule
\end{tabular}
\caption{Network architecture of DRMAC}
\label{tab:netarch}
\end{table}

\subsection{Hyper-parameters}
\begin{table}[h]
\centering
\begin{tabular}{ll}
  \toprule
  \textbf{Hyper-Parameters}  & \\
  \midrule
  epsilon start               & 1.0 \\
  epsilon finish             & 0.05 \\
  epsilon anneal time        & 50000 \\
  buffer size                & 5000 \\
  target update interval     & 200 \\
  hidden dimension for mixing network           & 32 \\
  $\beta$                 & 0.01 \\
  $\lambda$  &  0.0051\\

  \bottomrule
\end{tabular}
\caption{Hyper-Parameters of DRMAC}
\label{tab:hyparam}
\end{table}

\section{Experimental Setup}
\label{app:experimentalsetup}
All the Result are reported by averaging the results of 5 random seeds. The four test environments of the experiment are described as follows:

\textbf{Hallway}. This task revolves around multiple Markov chains where $n$ are initially distributed randomly across $n$ chains with varying lengths. The goal is for all agents to simultaneously reach the goal state, despite the constraint of partial observability. For increased challenge, $n$ is set to 4, and each chain has a unique length specified as (4, 6, 8, 10). At each time step, an agent has a limited observation of its current position, and it can choose from three actions: move left, move right, or remain stationary. An episode concludes when any agent reaches state $g$. The agents collectively succeed and receive a shared reward of 1 only if they all reach state $g$ concurrently. Otherwise, the reward is 0.

\textbf{Hallwaygroup}. This is a variant of Hallway which agents are divided into different groups and different groups have to arrive at different times. To intensify the challenge, we escalated the complexity by increasing both the number of agents and the lengths of the Markov chains. Specifically, in the Hallway benchmark, we set the number of agents to 4, with Markov chain lengths varying as (4,6,8,10). In the Hallwaygroup variant, we increased the number of agents to 7, dividing them into two groups. The lengths of the Markov chains for these two groups were set to (3,5,7) and (4,6,8,10), respectively.

\textbf{StarCraft Multi-Agent Challenge (SMAC)}.
This task revolves around a series of complex scenarios inspired by StarCraft\uppercase\expandafter{\romannumeral2}, a real-time strategy game. Decentralized agents engage in combat against the built-in AI, each having a limited field of vision restricted to adjacent units. Observations include relative positions, distances, unit types, and health statuses. Agents struggle to perceive the status of entities beyond their immediate vicinity, creating uncertainty. The action space varies across scenarios, often including move, attack, stop, and no-option. During training, global states with coordinates and features of all agents are accessible. Rewards are based on factors like damage infliction, eliminating units, or victory.

\textbf{SMAC-Communication}. To emphasize the role of communication, we select three super hard maps and further adopt the configuration used in \cite{ndq}. 
The specifics of the chosen scenarios are delineated as follows. 

\textbf{$5z\_vs\_1ul$}. A team of 5 Zealots faces a formidable Ultralisk. Victory requires mastering a complex micro-management technique involving positioning and attack timing.

\textbf{$1o\_10b\_vs\_1r$}. In a cliff-dominated terrain, an Overseer spots a Roach. 10 Banelings aim to eliminate the Roach for victory. Banelings can choose silence, relying on the Overseer to communicate its location, testing communication strategy efficiency.

\textbf{$1o\_2r\_vs\_4r$}. An Overseer encounters 4 Reapers. Allied units, 2 Roaches, must locate and eliminate the Reapers. Only the Overseer knows the Reapers' location, requiring effective communication for success.

\end{document}